# Analysis of Modern Computer Vision Models for Blood Cell Classification

Ryan Kim
Fremd High School
ryankim17920@gmail.com

Alexander Kim
University of Illinois Urbana-Champaign
ak98@illinois.edu

## Abstract

The accurate classification of white blood cells and related blood components is crucial for medical diagnoses. While traditional manual examinations and automated hematology analyzers have been widely used, they are often slow and prone to errors. Recent advancements in deep learning have shown promise for addressing these limitations. Earlier studies have demonstrated the viability of convolutional neural networks such as DenseNet, ResNet, and VGGNet for this task. Building on these foundations, our work employs more recent and efficient models to achieve rapid and accurate results. Specifically, this study used state-of-the-art architectures, including MaxVit, EfficientVit, EfficientNet, EfficientNetV2, and MobileNetV3. This study aimed to evaluate the performance of these models in WBC classification, potentially offering a more efficient and reliable alternative to current methods. Our approach not only addresses the speed and accuracy concerns of traditional techniques but also explores the applicability of innovative deep learning models in hematological analysis.

## Introduction

Blood is a vital fluid that transports oxygen and nutrients to the body's tissues and organs while also removing waste products[1]. It consists of plasma and various cellular components including red blood cells (RBCs), white blood cells (WBCs), and platelets[2]. WBCs, which make up approximately 1% of blood volume[2], play a crucial role in the body's immune system by defending against foreign pathogens and infections[3, 4].

There are five main types of WBCs: neutrophils, eosinophils, basophils, lymphocytes, and monocytes[5–7]. These can be categorized into granulocytes (neutrophils, eosinophils, and basophils) and agranulocytes (lymphocytes and monocytes) based on the presence or absence of [5, 6]. Additionally, erythroblasts in the blood are precursor cells in erythropoiesis, the production of red blood cells. They undergo a series of changes, including enucleation, to become mature red blood cells[8]. Platelets are also essential in the blood by aiding in clotting to prevent excessive bleeding. Immature granulocytes (IG) include promyelocytes, myelocytes, and metamyelocytes and are undeveloped WBCs that are expelled from the blood marrow[9]. Accurate identification and quantification of these WBC types are important for diagnosing various diseases and conditions, such as leukemia, immune disorders, and cancers[1, 10].

Traditional methods for WBC classification involve manual examination of stained blood smears under a microscope, which can be time consuming, subjective, and prone to errors[3]. Automated hematology analyzers can also be used; however, they lack the ability to utilize morphological information and cannot digitally preserve blood smears[4].

In recent years, deep learning (DL) techniques, particularly convolutional neural networks (CNNs), have shown promise for automating the process of WBC classification from digital blood smear images. These methods can automatically learn relevant features from images, potentially improving accuracy and efficiency compared to traditional approaches.

Earlier research has explored various CNN architectures for WBC classification. Tamang et al.[1] tested AlexNet[11], DenseNet[12], ResNet[13], VGGNet[14], and SqueezeNet[15] on a similar dataset and achieved high accuracy rates. DenseNet provided the greatest accuracy of 100% while ResNet achieved 99.91% accuracy. This was likely due to the highly connected nature of the layers through residual layers and dense connections in these models. Similarly, Asghar et al. [9] applies VGGNet, ResNet, InceptionV3[16], DenseNets, and MobileNetV2[17] and created their own 22-layer CNN model with 98.5% validation accuracy. Acevedo et al.[18], who also created the dataset used in this paper, used VGGNet and InceptionV3 to obtain 96% accuracies. Almezhghwi and Serte[3] also applied VGGNet, ResNets, and DenseNets but additionally utilized a generative adversarial network (GAN) for data augmentation. They also noted that the pretrained models performed slightly better in accuracy compared to the same models trained from scratch. Like other studies, DenseNets and ResNets obtained the highest accuracies of 98.8% accuracy and 97.4%, respectively for the largest model sizes. Ucar[19] utilizes a ShuffleNet[20] based model, which is another pre-trained model, to achieve 97% accuracy. Chen et al.[4] combines ResNet and DenseNet using a SCAM mechanism with a spatial attention module and a channel attention module.

Vision Transformers[21], which process images as a sequence of patches using transformer architectures[22], have also attracted interest in vision tasks. Ali etc. al[23] applied DenseNets, InceptionNet[24], ResNet, and Vision Transformer[21] to a similar blood dataset and found that the Vision Transformer obtained the highest validation accuracy. This demonstrates the potential of transformer-based architectures in WBC classification tasks.

While DenseNet, InceptionNet, ResNet, VGGNet, SqueezeNet, and AlexNet, developed in the mid-2010s, have been widely used in earlier studies, several newer models have since appeared as state-of-the-art in computer vision tasks. The aim of this study is to investigate and compare MaxVit, EfficientVit, EfficientNet, EfficientNetV2, and MobileNetV3. These are recent state-of-the-art models in terms of speed and/or accuracy. We utilize strong data augmentation to improve the generalization of models.

# Methods

## Dataset

The PBC dataset[25] comprises 17,092 professionally annotated images of normal blood cells from individuals without infection, hematologic or oncologic disease, and free from any pharmacologic treatment. Images were collected using the CellaVision DM96 analyzer at the Core Laboratory of the Hospital Clinic of Barcelona. Each image was 360 px x 363 px. jpg file. The distribution of the image data varied among the cell types, as shown in **Figure 1**.

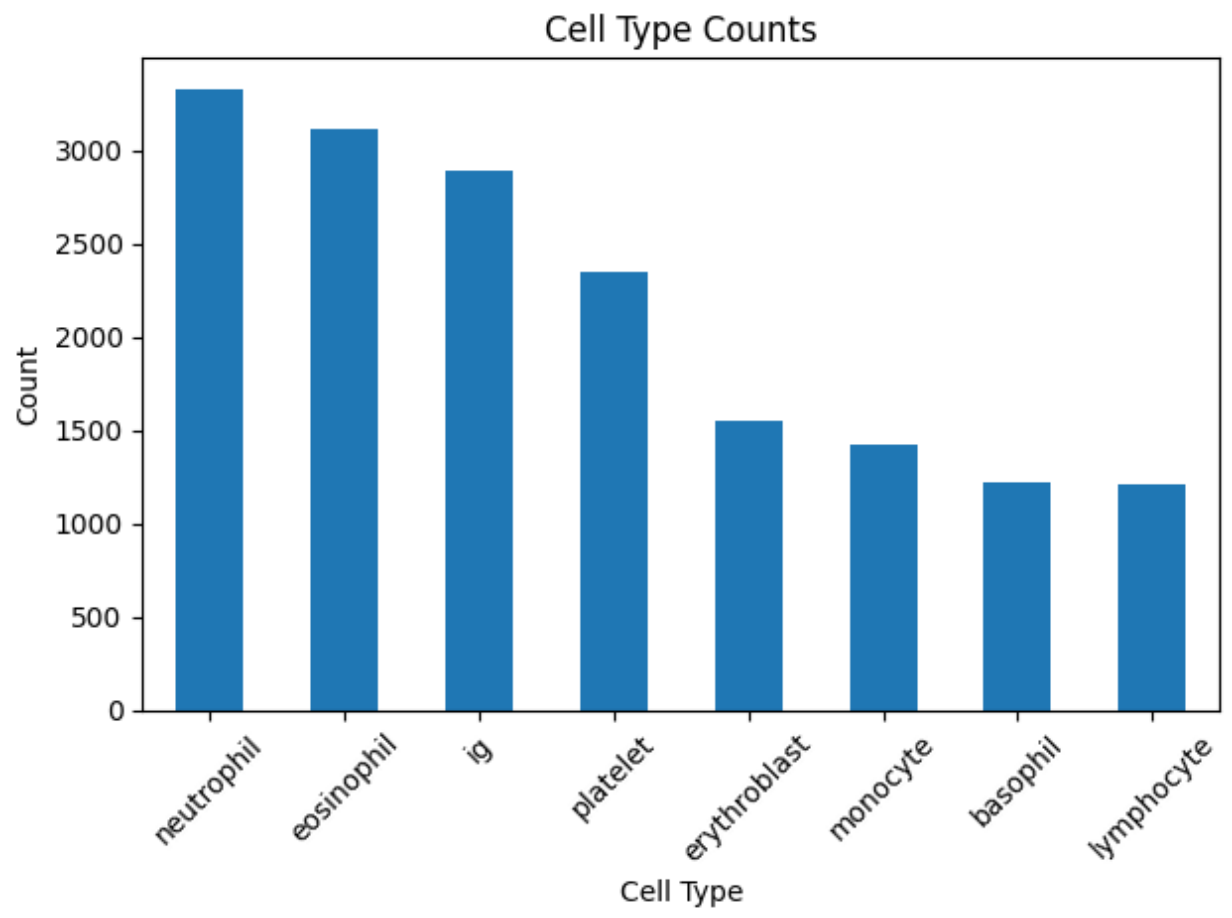

**Figure 1: Distribution of Images by Cell Type**

## Pre-Processing

Various techniques have been employed to process the PBC dataset, including image augmentation, reshaping for models, and data balancing. Image augmentation applies small random changes to the image data, which creates a more diverse dataset and improves the accuracy and generalization of models[26, 27]. Various augmentations were applied, as illustrated in **Figure 2**. Augmentations such as rotation, flips, shear, and translation were used to enhance the classification rigor of cells captured at different angles, perspectives, and locations. Color jitter was applied to address small color differences between images, further improving accuracy[28]. Special reshaping was applied to vision transformers, which required a rigid input shape for patching based on their pre-trained amount (see Model Selection and Transfer Learning for more details). The other CNN based models were reshaped to 360 px x 360 px. image shapes. Class weights for random sampling were also applied to address the uneven distribution of images so that all classes would be sampled evenly, effectively reducing bias[29].

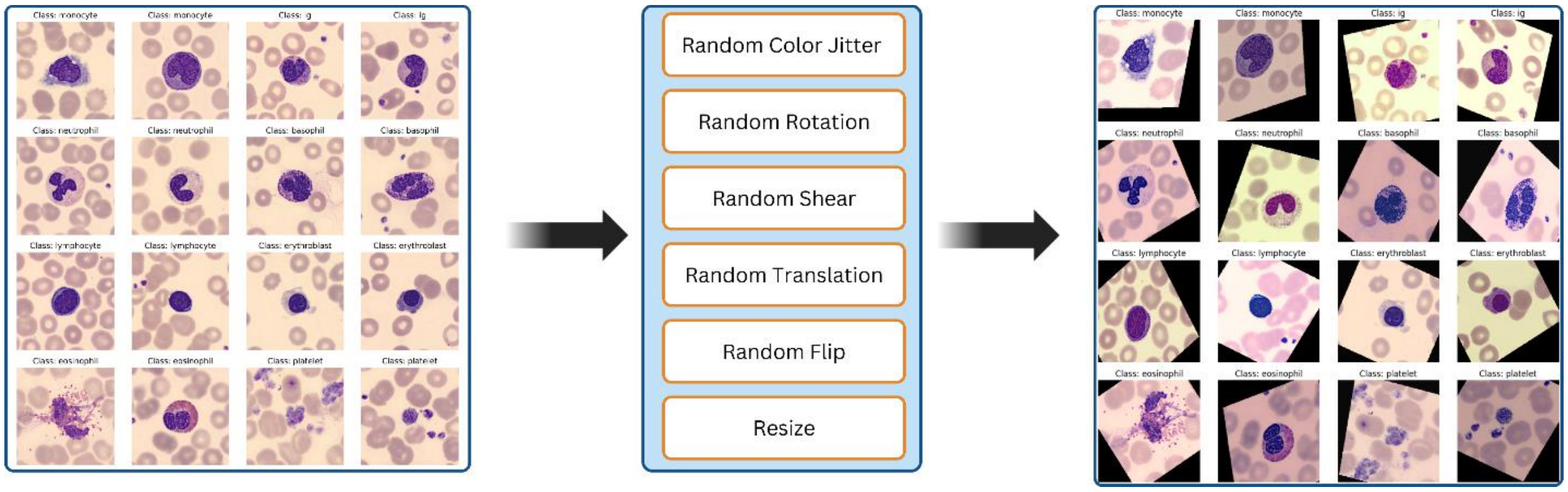

**Figure 2: Data Augmentation Process**

## Model Selection and Transfer Learning

Transfer learning, which involves fine-tuning a pre-trained model, is an effective way to improve model accuracy while reducing training time. These models are usually trained on a large dataset to capture generic patterns between images, enabling their use in other image-based tasks[30]. All models used in this study were pre-trained on ImageNet-1K[31], which contains over a million images across 1,000 categories. The performance statistics mentioned in this study (top-1 and top-5 accuracy) refer to the models' performance on the ImageNet validation set, providing a standardized benchmark for comparison.

GFLOPS, or Giga Floating Point Operations Per Second, is a measure of computational performance. It represents the number of billion floating-point operations a model can perform in one second, giving us an indication of the model's computational complexity and efficiency. A lower GFLOPS generally indicates faster inference times and lower computational requirements.

Model parameters are another crucial aspect to consider when evaluating deep learning models. The number of parameters in a model refers to the total number of learnable weights and biases in the neural network. These parameters define the model's capacity to learn and represent complex patterns in data.

In this study, we explored a range of pretrained models, focusing on the smallest variants (except for MobileNet) to balance performance and efficiency. Each model's ImageNet accuracy, GFLOPS, and number of parameters are provided to offer a comprehensive view of their capabilities, computational demands, and model complexity.

***Convolution Neural Networks***

- **MobileNetV3**: Optimized for mobile CPUs through simplified non-linear activations, reduced size, and depth[32], and the network architecture search (NAS) algorithm - Netadapt[33]. Both large and small models were used in the experiments. We evaluated both large and small variants, with the small model achieving 67.668% top-1 and 87.402% top-5 accuracy at just 0.06 GFLOPs and 2.5 million parameters, while the large model reached 75.274% top-1 and 92.566% top-5 accuracy with 0.22 GFLOPS and 5.5M parameters.

- **EfficientNet**: This model scales up CNNs by balancing the depth, width, and resolution to achieve optimal accuracy while being smaller and faster[34]. Known for its high accuracy and speed, EfficientNet_B0 achieved 77.692% top-1 and 93.532% top-5 accuracy with 0.39 GFLOPs and 5.3M parameters, making it a popular choice in the field.
- **EfficientNetv2**: An improved version of EfficientNet with higher accuracy and speed, created through enhanced neural architecture search[35] and optimizations, such as fused MBConv layers[36]. It outperforms its predecessor with 84.228% top-1 and 96.878% top-5 accuracy with 8.37 GFLOPs and 21.5M parameters with the EfficientNetv2 small model, demonstrating considerable progress in CNN design.

***Transformer Models***

- **MaxViT**: It features a scalable and efficient multi-axis attention layer that combines local and global attention with convolutions. This architecture achieves high performance in various vision tasks with linear complexity[37]. It demonstrates impressive performance across various vision tasks, reaching 83.7% top-1 and 96.722% top-5 accuracy at 5.56 GFLOPs and 30.9M parameters with its tiny model. To accommodate MaxViT's input requirements, the images were reshaped to 224 x 224 pixels.
- **EfficientVit**: Utilizes ReLU linear attention instead of SoftMax attention, which provides up to 3.3-4.5 times speedup for the multi-head attention layers. Additionally, the model uses convolution layers for earlier stages instead of fully attention-based systems[38]. EfficientVit-B1 uses a resolution of 288 x 288 pixels and achieves 80.410% top-1 and 94.984% top-5 with 1.72 GFLOPS and 9.1M parameters.

## Metrics

Model evaluation metrics included True Positive (TP), True Negative (TN), False Positive (FP), and False Negative (FN) counts. While accuracy is a common metric, the F1-score is particularly useful for unbalanced datasets, such as the one used in this study[39].

- **Accuracy**: $\frac{\mathrm{TP} + \mathrm{TN}}{\mathrm{TP} + \mathrm{TN} + \mathrm{FP} + \mathrm{FN}}$
- **Precision**: $\frac{\mathrm{TP}}{\mathrm{TP} + \mathrm{FP}}$
- **Recall**: $\frac{\mathrm{TP}}{\mathrm{TP} + \mathrm{TN}}$
- **F1-Score**: $\frac{2 \times \mathrm{Precision} \times \mathrm{Recall}}{\mathrm{Precision} + \mathrm{Recall}}$

### Training

Models were trained using a relative batch size of 2048, achieved through a batch size of 64, with gradient accumulation over 32 steps. They additionally used the Adam optimizer[40] with learning rates of 0.0005 for all. To assess model performance, we employed a stratified 5-fold cross-validation method. The dataset was divided into five stratified folds, with each fold serving as a validation set for one iteration of model training. To accelerate the training process, we implemented mixed precision training[41]. The models were trained for a maximum of 15 epochs, with an early stopping mechanism in place. This mechanism would halt training if no improvement in validation accuracy was observed for 25 consecutive validation checks, with each check occurring after every gradient accumulation step.

## Results

Models were evaluated on the metrics by aggregating validation sets in the stratified 5-folds.

| MODEL | METRICS |
|---|---|

MOBILENETV3 – SMALL

| | basophil | eosinophil | erythroblast | ig | lymphocyte | monocyte | neutrophil | platelet | macro avg | weighted avg/ accuracy |
|---|---|---|---|---|---|---|---|---|---|---|
| precision | 0.9735 | 0.9997 | 0.9758 | 0.9700 | 0.9819 | 0.9747 | 0.9816 | 0.9966 | 0.9817 | 0.9833 |
| recall | 0.9943 | 0.9958 | 0.9878 | 0.9603 | 0.9811 | 0.9775 | 0.9784 | 0.9983 | 0.9842 | 0.9833 |
| f1-score | 0.9838 | 0.9978 | 0.9817 | 0.9651 | 0.9815 | 0.9761 | 0.9800 | 0.9974 | 0.9829 | 0.9833 |
| count | 1218 | 3117 | 1551 | 2895 | 1214 | 1420 | 3329 | 2348 | 17092 | 17092 |

MOBILENETV3 – LARGE

| | basophil | eosinophil | erythroblast | ig | lymphocyte | monocyte | neutrophil | platelet | macro avg | weighted avg/ accuracy |
|---|---|---|---|---|---|---|---|---|---|---|
| precision | 0.9942 | 0.9990 | 0.9929 | 0.9791 | 0.9942 | 0.9697 | 0.9838 | 0.9979 | 0.9889 | 0.9888 |
| recall | 0.9926 | 0.9990 | 0.9897 | 0.9713 | 0.9942 | 0.9915 | 0.9826 | 0.9987 | 0.9900 | 0.9888 |
| f1-score | 0.9934 | 0.9990 | 0.9913 | 0.9752 | 0.9942 | 0.9805 | 0.9832 | 0.9983 | 0.9894 | 0.9888 |
| count | 1218 | 3117 | 1551 | 2895 | 1214 | 1420 | 3329 | 2348 | 17092 | 17092 |

EFFICIENTNET _B0

| | basophil | eosinophil | erythroblast | ig | lymphocyte | monocyte | neutrophil | platelet | macro avg | weighted avg/ accuracy |
|---|---|---|---|---|---|---|---|---|---|---|
| precision | 0.9886 | 0.9994 | 0.9898 | 0.9830 | 0.9845 | 0.9887 | 0.9876 | 0.9991 | 0.9901 | 0.9907 |
| recall | 0.9967 | 1.000 | 0.9974 | 0.9765 | 0.9942 | 0.9866 | 0.9838 | 0.9987 | 0.9918 | 0.9907 |
| f1-score | 0.9926 | 0.9997 | 0.9936 | 0.9797 | 0.9893 | 0.9877 | 0.9857 | 0.9989 | 0.9909 | 0.9907 |
| support | 1218 | 3117 | 1551 | 2895 | 1214 | 1420 | 3329 | 2348 | 17092 | 17092 |

EFFICIENTNET V2-SMALL

| | basophil | eosinophil | erythroblast | ig | lymphocyte | monocyte | neutrophil | platelet | macro avg | weighted avg/ accuracy |
|---|---|---|---|---|---|---|---|---|---|---|
| precision | 0.9943 | 0.9994 | 0.9955 | 0.9803 | 0.9942 | 0.9874 | 0.9879 | 0.9996 | 0.9923 | 0.9919 |
| recall | 0.9967 | 0.9990 | 0.9974 | 0.9817 | 0.9934 | 0.9915 | 0.9847 | 0.9983 | 0.9929 | 0.9919 |
| f1-score | 0.9955 | 0.9992 | 0.9965 | 0.9810 | 0.9938 | 0.9895 | 0.9863 | 0.9989 | 0.9926 | 0.9919 |
| support | 1218 | 3117 | 1551 | 2895 | 1214 | 1420 | 3329 | 2348 | 17092 | 17092 |

MAXVIT_TINY

| | basophil | eosinophil | erythroblast | ig | lymphocyte | monocyte | neutrophil | platelet | macro avg | weighted avg/ accuracy |
|---|---|---|---|---|---|---|---|---|---|---|
| precision | 0.9846 | 0.9990 | 0.9923 | 0.9823 | 0.9934 | 0.9929 | 0.9853 | 0.9996 | 0.9912 | 0.9910 |
| recall | 0.9959 | 0.9997 | 0.9961 | 0.9751 | 0.9959 | 0.9887 | 0.9871 | 0.9979 | 0.9921 | 0.9910 |
| f1-score | 0.9902 | 0.9994 | 0.9942 | 0.9787 | 0.9947 | 0.9908 | 0.9862 | 0.9987 | 0.9916 | 0.9910 |
| support | 1218 | 3117 | 1551 | 2895 | 1214 | 1420 | 3329 | 2348 | 17092 | 17092 |

| EFFICIENTVIT-B1 | | basophil | eosinophil | erythroblast | ig | lymphocyte | monocyte | neutrophil | platelet | macro avg | weighted avg/ accuracy |
|---|---|---|---|---|---|---|---|---|---|---|---|
| | precision | 0.9878 | 0.9990 | 0.9917 | 0.9786 | 0.9893 | 0.9908 | 0.9876 | 0.9991 | 0.9905 | 0.9905 |
| | recall | 0.9967 | 0.9984 | 0.9968 | 0.9810 | 0.9918 | 0.9866 | 0.9823 | 0.9979 | 0.9914 | 0.9905 |
| | f1-score | 0.9922 | 0.9987 | 0.9942 | 0.9798 | 0.9905 | 0.9887 | 0.9849 | 0.9985 | 0.9910 | 0.9905 |
| | support | 1218 | 3117 | 1551 | 2895 | 1214 | 1420 | 3329 | 2348 | 17092 | 17092 |

The models' predictions were also aggregated into confusion matrixes, as seen in Appendix.

# Discussion

## Overview

All the evaluated models demonstrated high accuracy and F1-scores, with the best-performing model, EfficientNetV2 Small, achieving an accuracy of 99.19% and an F1-score of 99.26%. However, increasing the number of FLOPs did not significantly enhance performance. For instance, MobileNetV3 Small achieved a validation accuracy of 98.33% and an F1-score of 98.29% with only 0.06 GFLOPs, whereas the top model required 8.37 GFLOPs—139.5 times more.

## Evaluation of Results

1. **Dataset Complexity and Model Capacity:** The simplicity of the dataset allows small models to effectively capture the necessary features, leading to high performance without requiring additional model complexity. This suggests that the dataset does not present enough challenges to justify the increased capacity of the larger models.

2. **Impact of Outliers:** Outliers in the dataset likely prevent any model from achieving an accuracy higher than 99.1%. These outliers may introduce noise, which hinders further performance improvements, even with more complex models.
3. **Data Preprocessing and Reshaping:** Resizing images for Vision Transformers might lead to loss of crucial information, negatively affecting their performance.
4. **Augmentation Strategies:** The current augmentation techniques may not be sufficiently robust to enhance model generalization. More aggressive or varied augmentations could help models learn more diverse features, improving their performance on unseen data.
5. **Comparison of Model Architectures:** Small models, such as MobileNetv3 Small, demonstrate impressive efficiency, achieving high accuracy with minimal computational resources. This is the fastest model tested for blood cell classification and shows the practicality of utilizing it for quick and accurate results.

# Conclusion

This study evaluated the performance of several state-of-the-art computer vision models for white blood cell classification, including MaxVit, EfficientNet, EfficientNetV2, and MobileNetV3. Our results demonstrated that these modern architectures could achieve high accuracy in classifying blood cell types, with even the smallest models reaching impressive performance levels.

Notably, MobileNetV3 Small achieved 98.33% validation accuracy with only 0.06 GFLOPs, highlighting its exceptional efficiency for this task. Lightweight models can be highly effective for blood cell classification, potentially enabling rapid and accurate analysis in resource-constrained environments.

Additionally, we observed that increasing the model complexity and computational requirements did not necessarily lead to significant improvements in classification accuracy. This indicates that the complexity of the dataset may not warrant the use of larger, more resource-intensive models.

Future work could focus on further optimizing these models for deployment on mobile or edge devices, exploring more advanced data augmentation techniques to enhance generalization, and investigating the performance of the models on more diverse and challenging datasets. Additionally, addressing the impact of outliers and potential information loss during image resizing can lead to even higher classification accuracies.

In conclusion, our findings demonstrate the viability of using efficient state-of-the-art computer vision models for accurate and rapid blood cell classification, paving the way for improved diagnostic tools in hematology.

# Appendix

Confusion matrices were created for all tested models, as shown below.

| MODEL | CONFUSION MATRIXES |
|---|---|

MOBILENETV3-SMALL

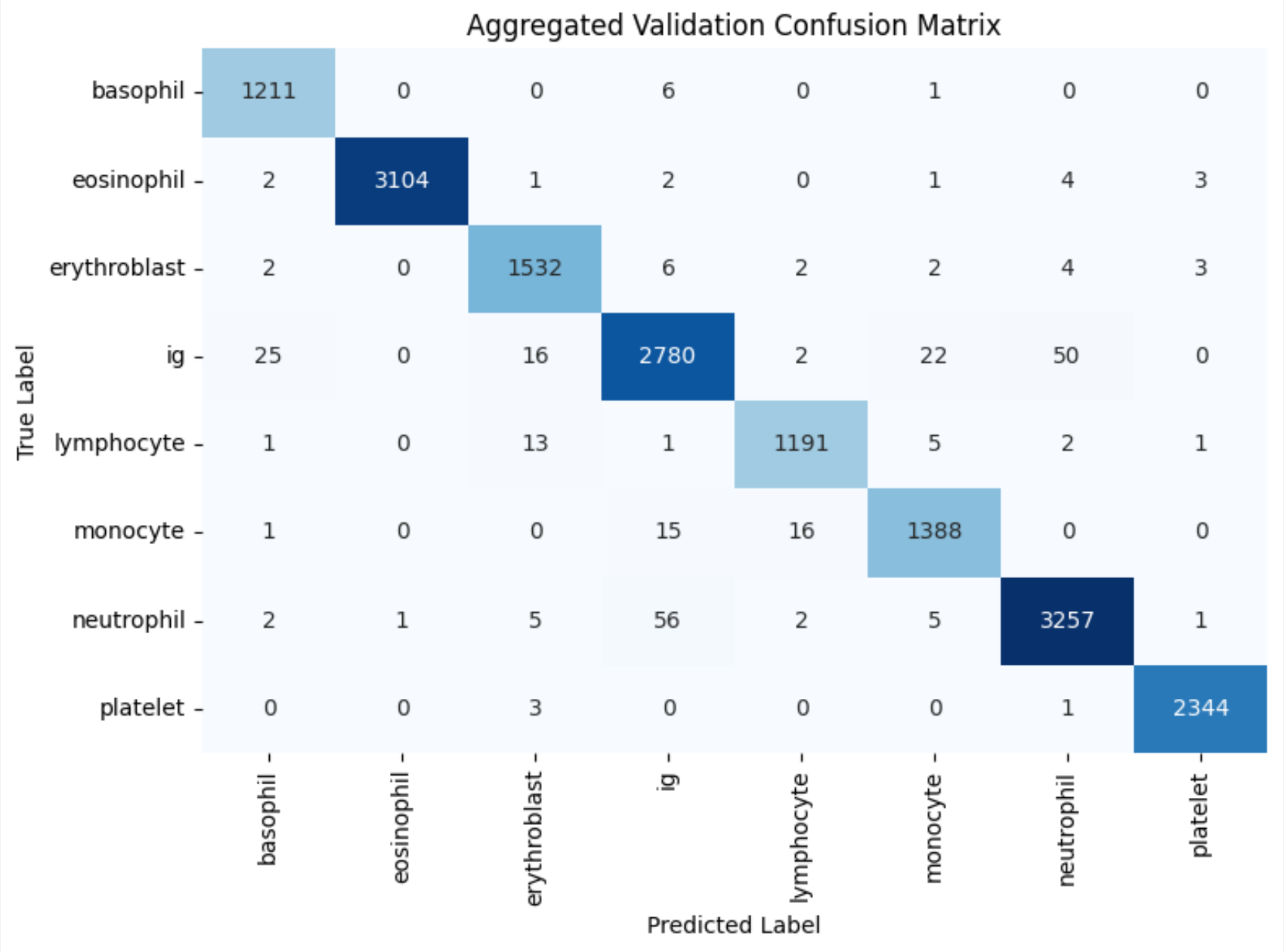

MOBILENETV3-LARGE

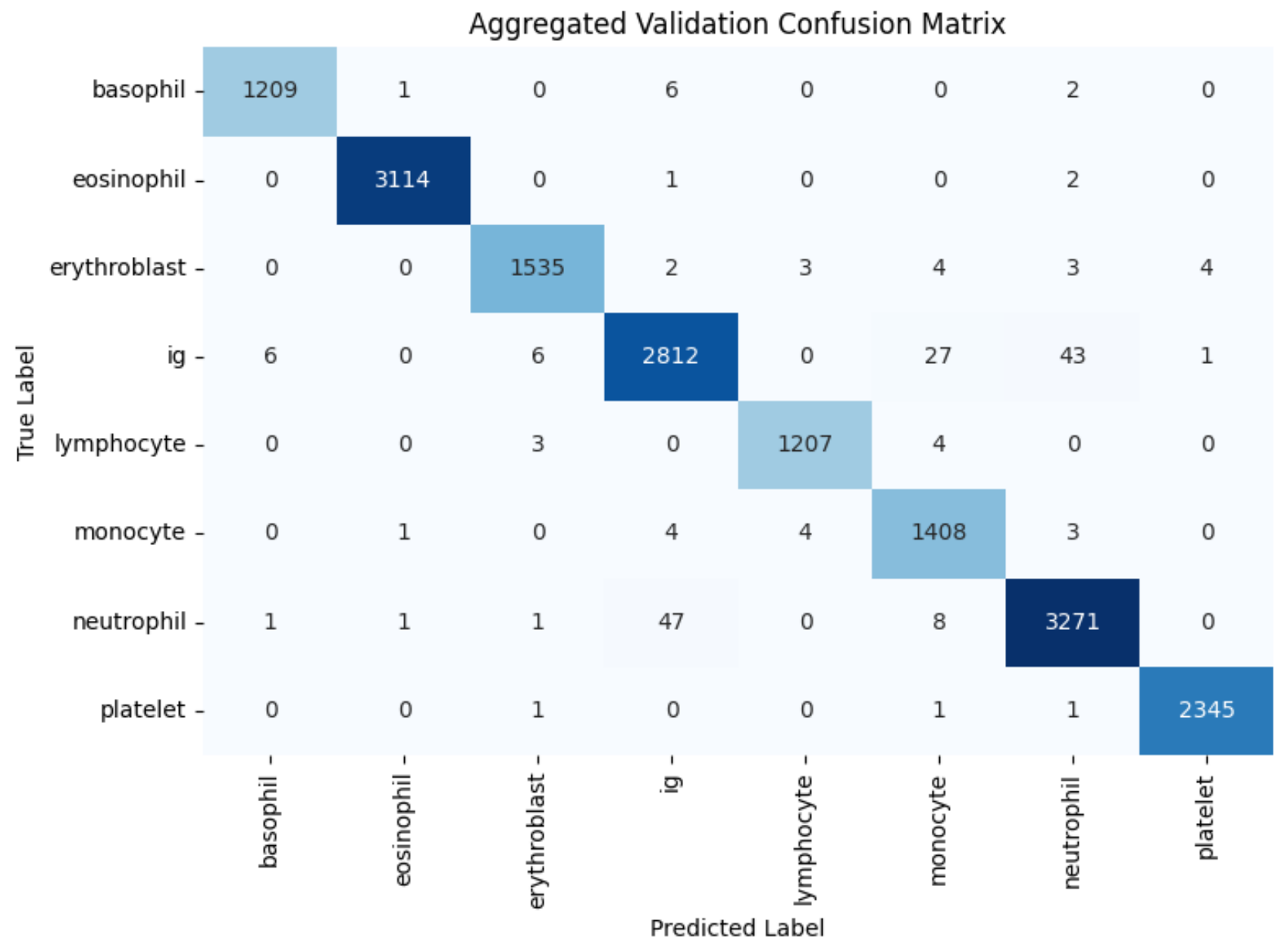

EFFICIENTNET_B0

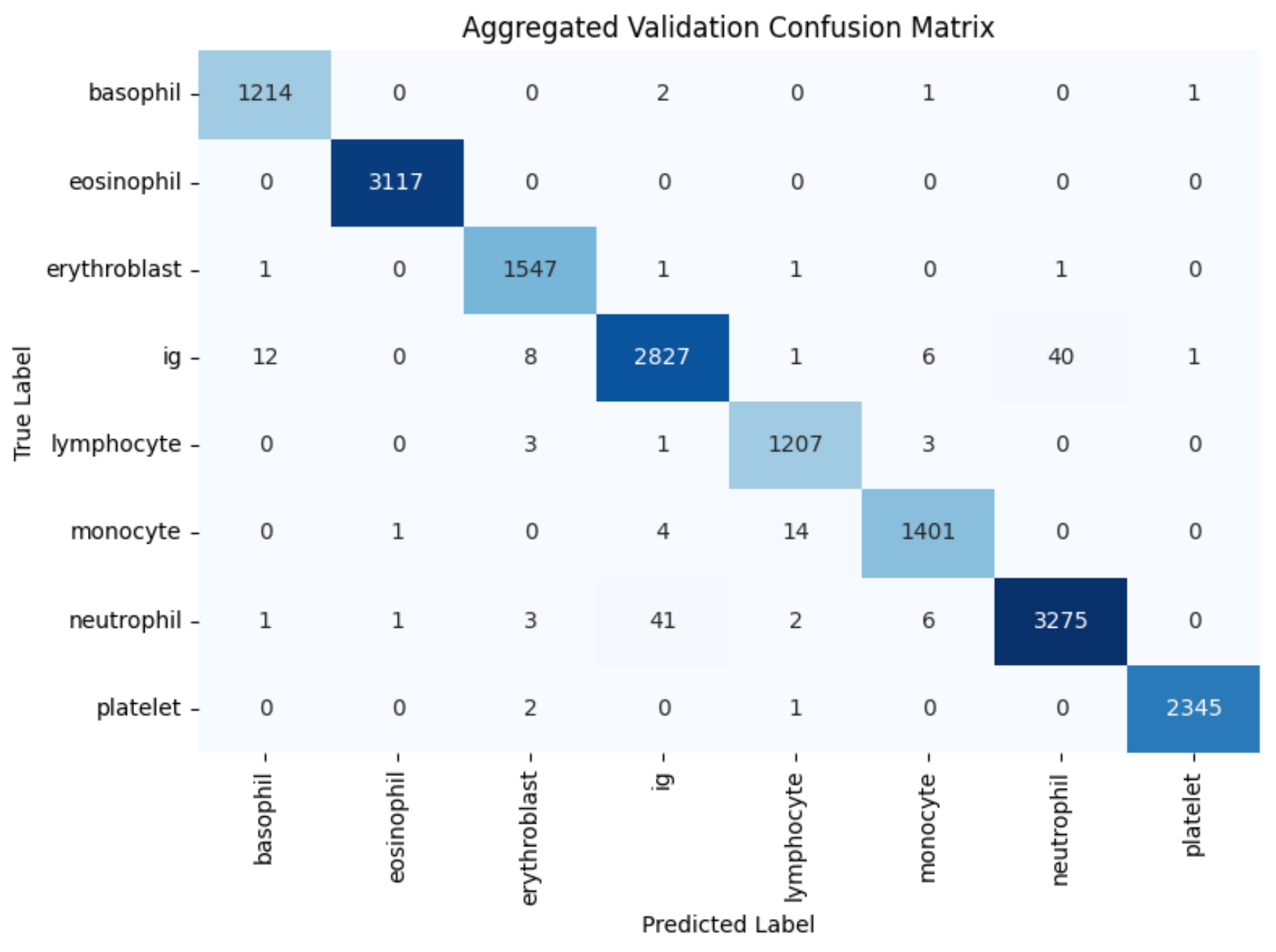

EFFICIENTNETV2-SMALL

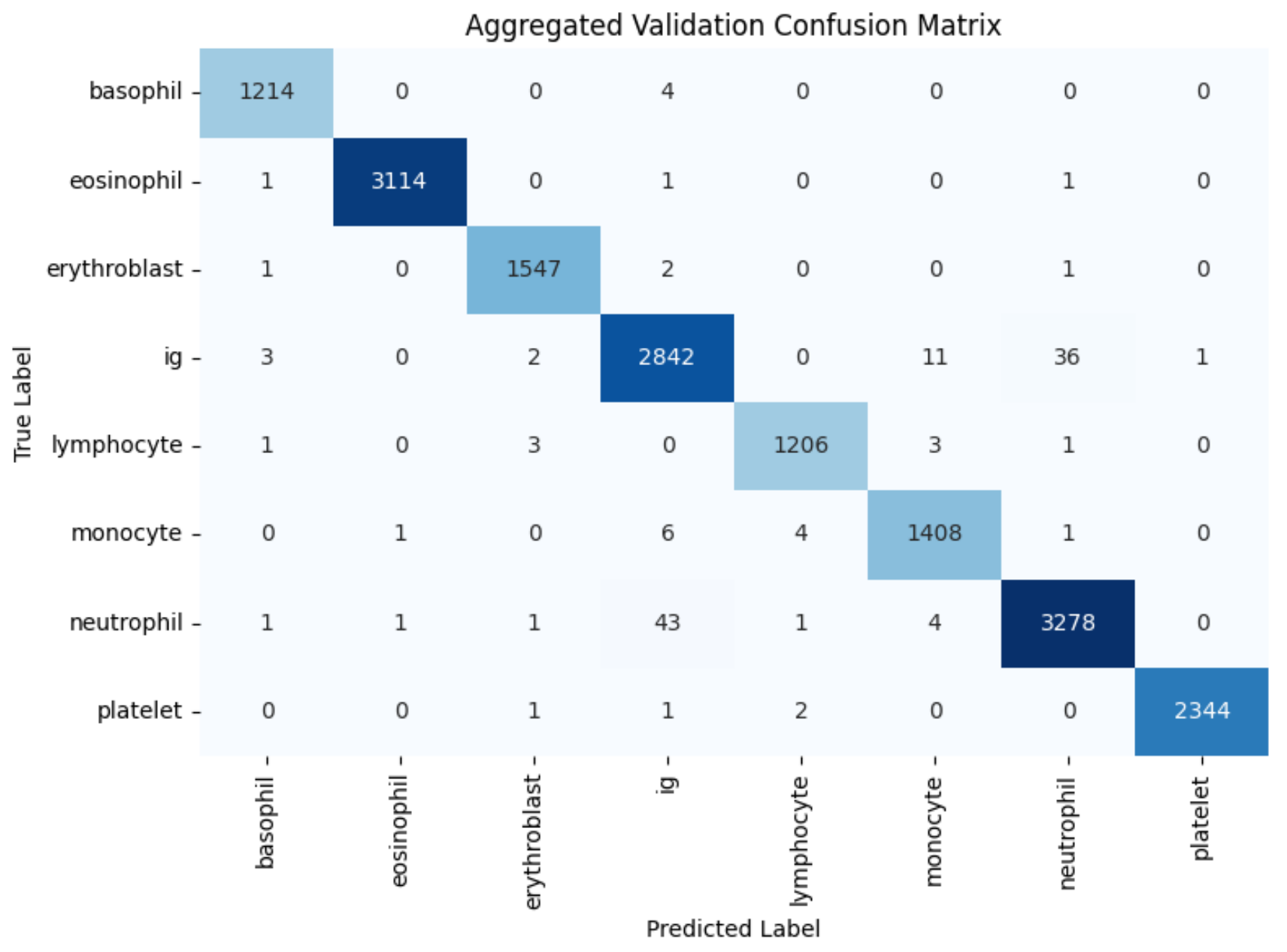

MAXVIT-TINY

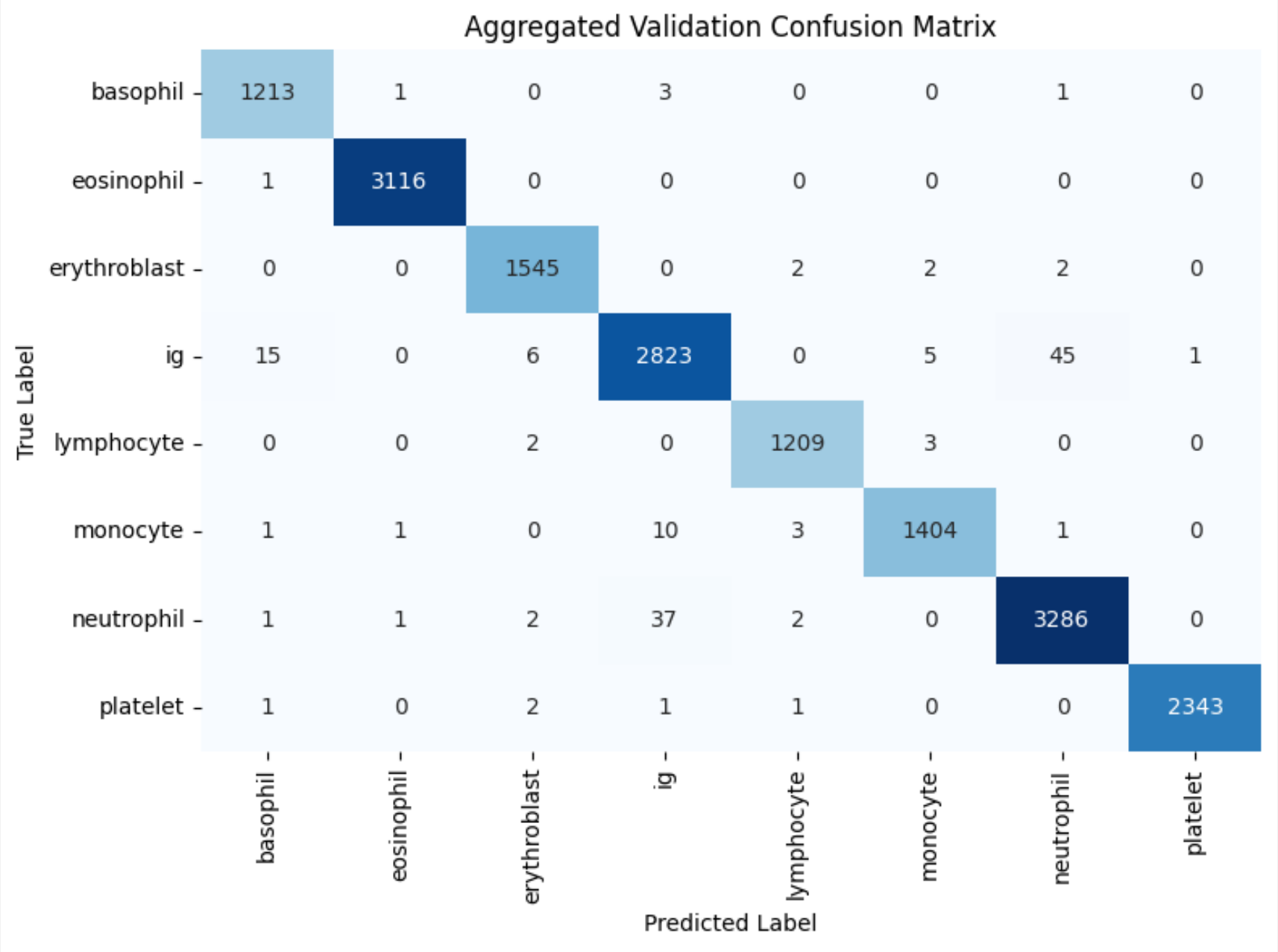

EFFICIENTVIT-B1

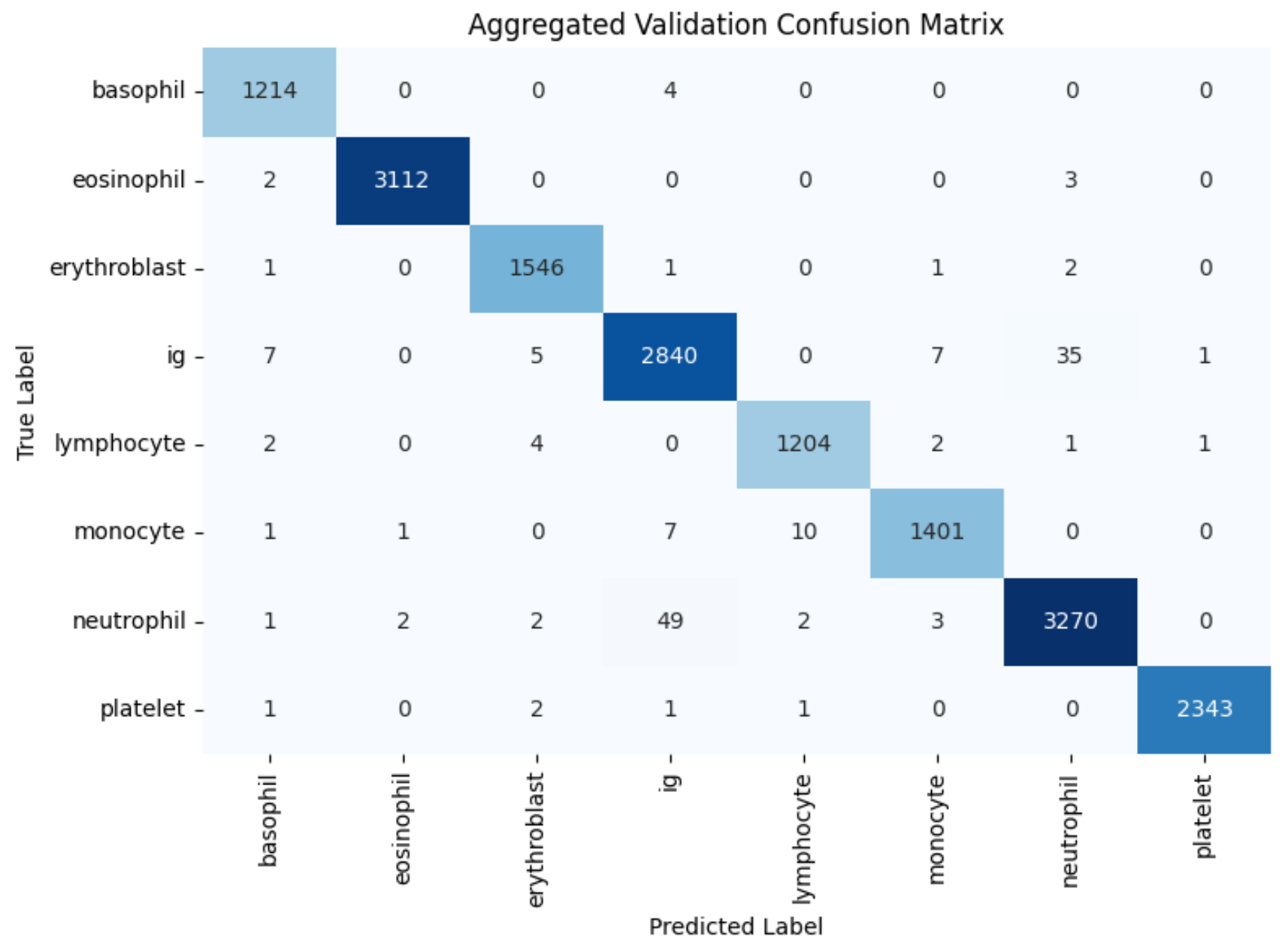